\documentclass[twocolumn,aps,floatfix]{revtex4}
\usepackage{latexsym,amsmath,amssymb,amsbsy}
\usepackage{graphicx}
\usepackage{nccfoots}
\usepackage[symbol]{footmisc}

\begin{document}


\title{Correlations of $\alpha$-particles in splitting of $^{12}$C nuclei by neutrons\\
 of energy of 14.1 MeV}


\author{\textbf{R.~R.~Kattabekov$^{\textbf{1), 2)}}$, K.~Z.~Mamatkulov$^{\textbf{1), 3)}}$, D.~A.~Artemenkov$^{\textbf{1)}}$, V.~Bradnova$^{\textbf{1)}}$, P.~I.~Zarubin$^{\textbf{1)~*}}$, I.~G.~Zarubina$^{\textbf{1)}}$, L.~Majling$^{\textbf{4)}}$, V.~V.~Rusakova$^{\textbf{1)}}$, A.~B.~Sadovsky$^{\textbf{1)}}$}
\Footnotetext{1)}{Joint Institute for Nuclear Research, Dubna, Moscow region, Russia}
\Footnotetext{2)}{Institute for Physics and Technology, Uzbek Academy of Sciences, Tashkent, Republic of Uzbekistan}
\Footnotetext{3)}{Dzhizak State Pedagogical Institute, Dzhizak, Republic of Uzbekistan}
\Footnotetext{4)}{Institute of Nuclear Physics, Rzez, Czech Republic}
\Footnotetext{*}{E-mail: \texttt{zarubin@lhe.jinr.ru}}}

\indent \par
\noindent \affiliation{Received XX, 2014}

\begin{abstract} 
Correlations of $\alpha$-particles are studied on statistics of 400 events of splitting $^{12}$C $\rightarrow$ 3$\alpha$ in nuclear track emulsion exposed to $14.1 MeV$ neutrons. The ranges and emission angles of the $\alpha$-particles are measured. Distributions over energy of $\alpha$-particle pairs and triples are obtained.\par

\end{abstract}

\maketitle

\indent Nuclear track emulsion (NTE) exposed to neutrons of energy of $14.1 MeV$ produced in a low energy reaction $d + t \rightarrow n + \alpha$ allows one to study the ensembles of triples of $\alpha$-particles produced in disintegrations of carbon nuclei of NTE composition. Energy transferred to $\alpha$-particles is sufficient to measure their ranges and directions and, at the same time, it remains below the thresholds of background channels. Such an approach to the experimental study emerged with the advent of neutron generators. Most completely the reaction $^{12}C(n,n')3\alpha$  was studied quite a long time ago [1]. An initial objective of this analysis was limited to $\alpha$-calibration of NTE, recently reproduced by \lq\lq TD Slavich\rq\rq [2]. A significant number of $\alpha$-triples of the reaction $^{12}C(n,n')3\alpha$ reached 1200 in a short time made it possible to analyze it on a large statistics as well as to create a commonly available bulk of experimental data. This bulk is useful for a direct comparison with the $\alpha$-cluster models of the $^{12}$C nucleus.\par
\indent A topical physical interest to the reaction $^{12}C(n,n')3\alpha$ is as follows. Information about a probability of presence of configurations of $\alpha$-particle clusters with different angular momenta is of fundamental importance for description of the structure of light nuclei. Studying the dissociation of relativistic nuclei $^{9}$Be in NTE [3, 4-6] the BECQUEREL Collaboration confirmed the two-body model of the nucleus $^{9}$Be in which is dominating the superposition of the neutron and an unstable $^{8}$Be nucleus in states with spin and parity $0^{+}$ and $2^{+}$ presenting with similar weights is dominant. In this way a basis appears for asking questions about the contributions $\alpha$-cluster configurations in the angular momenta of the ground states of heavier nuclei.\par
\indent Traditionally the nucleus $^{12}$C is regarded as a \lq\lq laboratory\rq\rq for the development the $\alpha$-particle clustering concepts. It is a possible that in the ground state of $^{12}$C$_{g.s.}$ there are two pairs of $\alpha$-clusters with orbital angular momenta equal to 2 (D-wave). In this case the basic configurations are $^{8}$Be nuclei in the first excited state 2$^{+}$. In a classical pattern one may imagine a rotation in opposite directions of two $\alpha$-clusters with angular momenta equal to 2 around a common center represented by a third $\alpha$-cluster. Then the remaining combination of two $\alpha$-clusters should correspond to the ground state of the nucleus $^{8}$Be with spin and parity $0^{+}$ (S-wave). As a result the superposition of the pair states in the ensemble of three $\alpha$-clusters leads to a zero spin in $^{12}$C$_{g.s.}$. Naturally, this simplified model requires a quantum-mechanical consideration. Nevertheless, its validity should be confirmed by an intensive formation of states $^{8}$Be$_{2+}$ and $^{8}$Be$_{g.s.}$ with a predominance of the former one in reactions of knocking of $\alpha$-particles from $^{12}$C nuclei.\par
\indent Such a concept does not contradict the mechanism of the synthesis of the nucleus $^{12}$C accepted in nuclear astrophysics. Fusion of a triple of $\alpha$-particles occurs through its second excited state $0^{+}_{2}$ (the Hoyle state) located on 270 keV above the breakup threshold $^{12}C \rightarrow 3\alpha$. Basically, each pair of $\alpha$-particles in it corresponds to $^{8}$Be$_{g.s.}$. In the transition $0^{+}_{2} \rightarrow 2^{+}_{1}$ with emission of a photon to the first excited state of $^{12}$C, which is bound one, an $\alpha$-pair in the D-wave should arise in a 3$\alpha$ ensemble in order to provide conservation of the angular momentum. The subsequent transition to $^{12}$C$_{g.s.}$, which is also accompanied by emission of a photon leads to the formation of another $\alpha$-particle pair in the D-wave state. This pair should have an opposite angular momentum with respect to the first pair to ensure zero spin value of the ground state $^{12}$C$_{g.s.}$. Thus, the nucleus $^{12}$C$_{g.s.}$ does acquire polarization. Figuratively being expressed it does conserve an \lq\lq invisible rotation\rq\rq.\par

\begin{figure}
\includegraphics[width=0.45\textwidth]{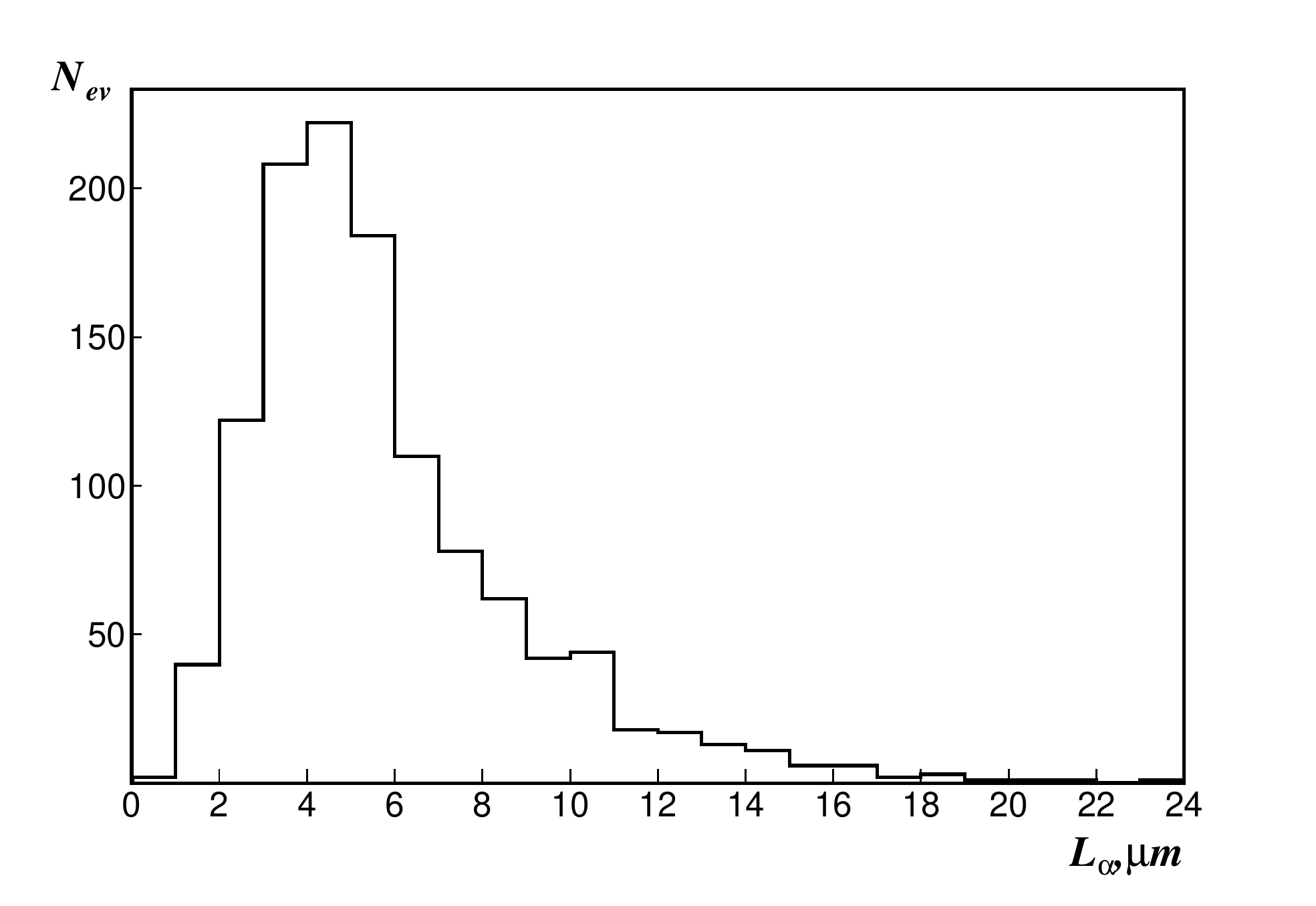}
\caption{Distribution of $\alpha$-particles over ranges $L_{\alpha}$.}
\label{fig:1}
\end{figure} 
\begin{figure}
\includegraphics[width=0.45\textwidth]{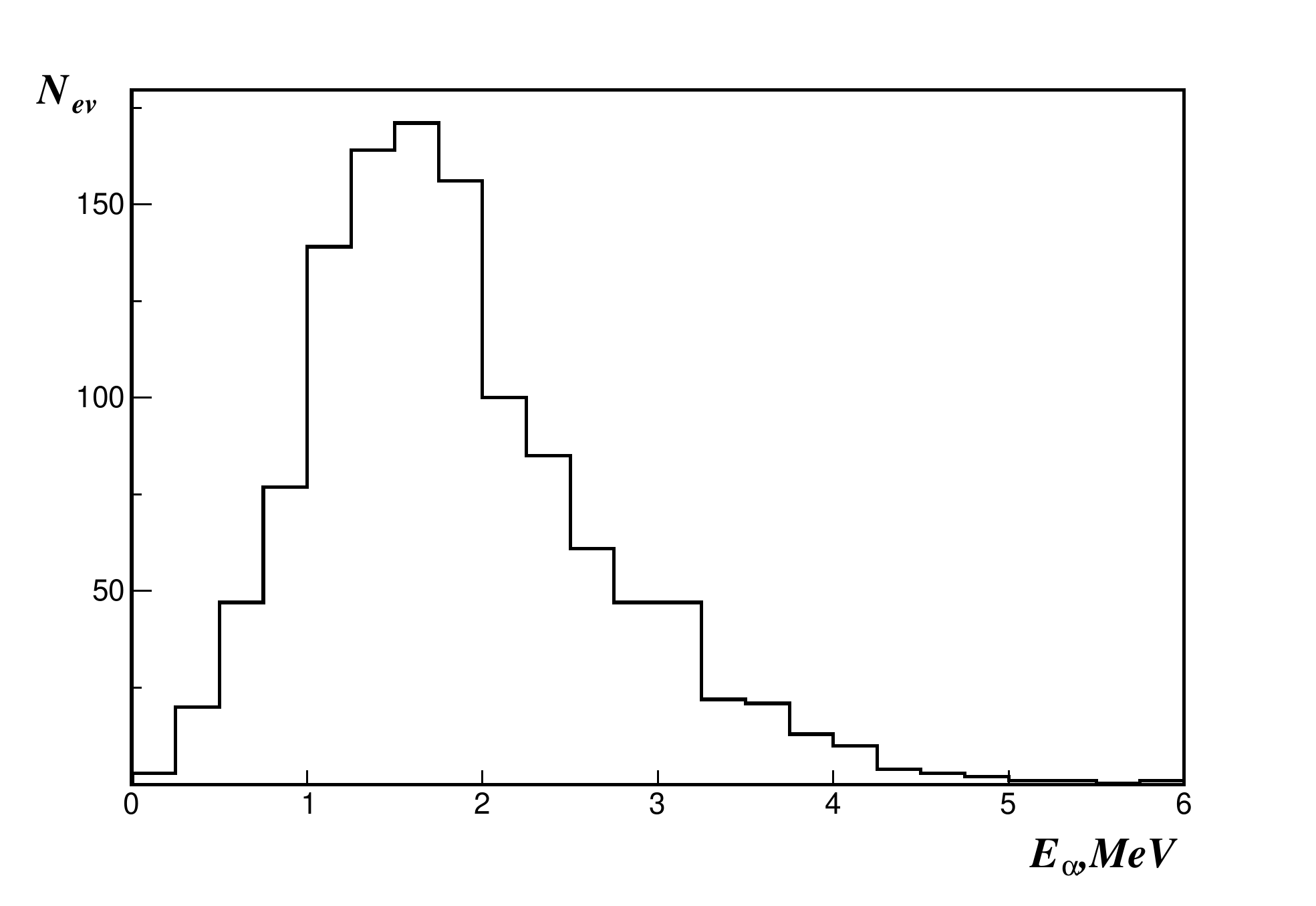}
\caption{Distribution of $\alpha$-particles over energy $E_{\alpha}$.}
\label{fig:2}
\end{figure} 

\indent The ratio of the yields of $\alpha$-particle pairs through the states $^{8}$Be$_{2+}$ and $^{8}$Be$_{g.s.}$ in disintegrations of nuclei $^{12}$C not accompanied by a transfer of the angular momentum is a key parameter which should reflect the spin-cluster structure $^{12}$C$_{g.s.}$. Analysis of interactions in NTE exposed to neutrons of energy near the threshold of the $^{12}$C splitting allows one to determine this and other characteristics of the reaction $^{12}C(n,n')3\alpha$.\par
\indent Exposure of NTE to neutrons of energy $14.1 MeV$ was performed on one of devices DVIN of an applied destination [7]. A neutron generator of the device provided a flow of $5 \times 10 ^{7}$ neutrons/s in the full solid angle. The NTE stack was placed on a top cover of the device DVIN approximately 10$cm$ above a tritium target. The stack consisted of several layers of NTE BR-2 of size of $9$ to $12 cm^{2}$ at thickness of 107 microns poured onto glass plates of thickness of 2$mm$. The neutron generator gave rise to an irreducible background of X-ray radiation. This background was detected by the NTE layers with decreasing brightness as the absorption in the glasses grows which allowed one to select layers with a low X-ray backlighting. NTE is comparable to a liquid hydrogen target on density of hydrogen. Therefore, the main background in NTE exposed to neutrons, is presented by recoil protons. Overlaying of tracks that would be imitating 3$\alpha$-disintegrations were reduced to a negligible level by choice of the exposure time of 40 min.\par

\begin{figure}
\includegraphics[width=0.45\textwidth]{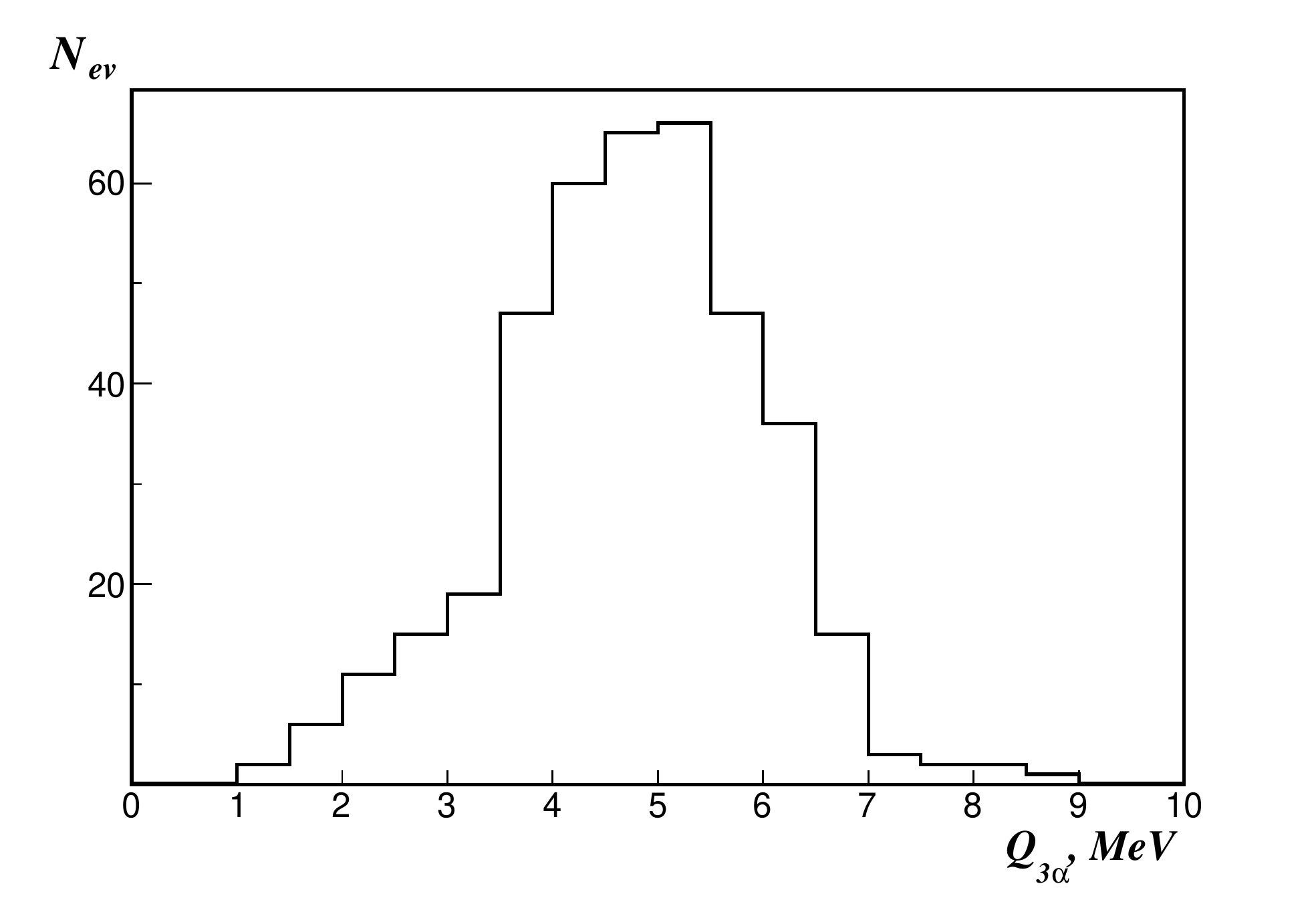}
\caption{Distribution  triples of $\alpha$-particles over energy $Q_{3\alpha}$.}
\label{fig:3}
\end{figure}
\begin{figure}
\includegraphics[width=0.45\textwidth]{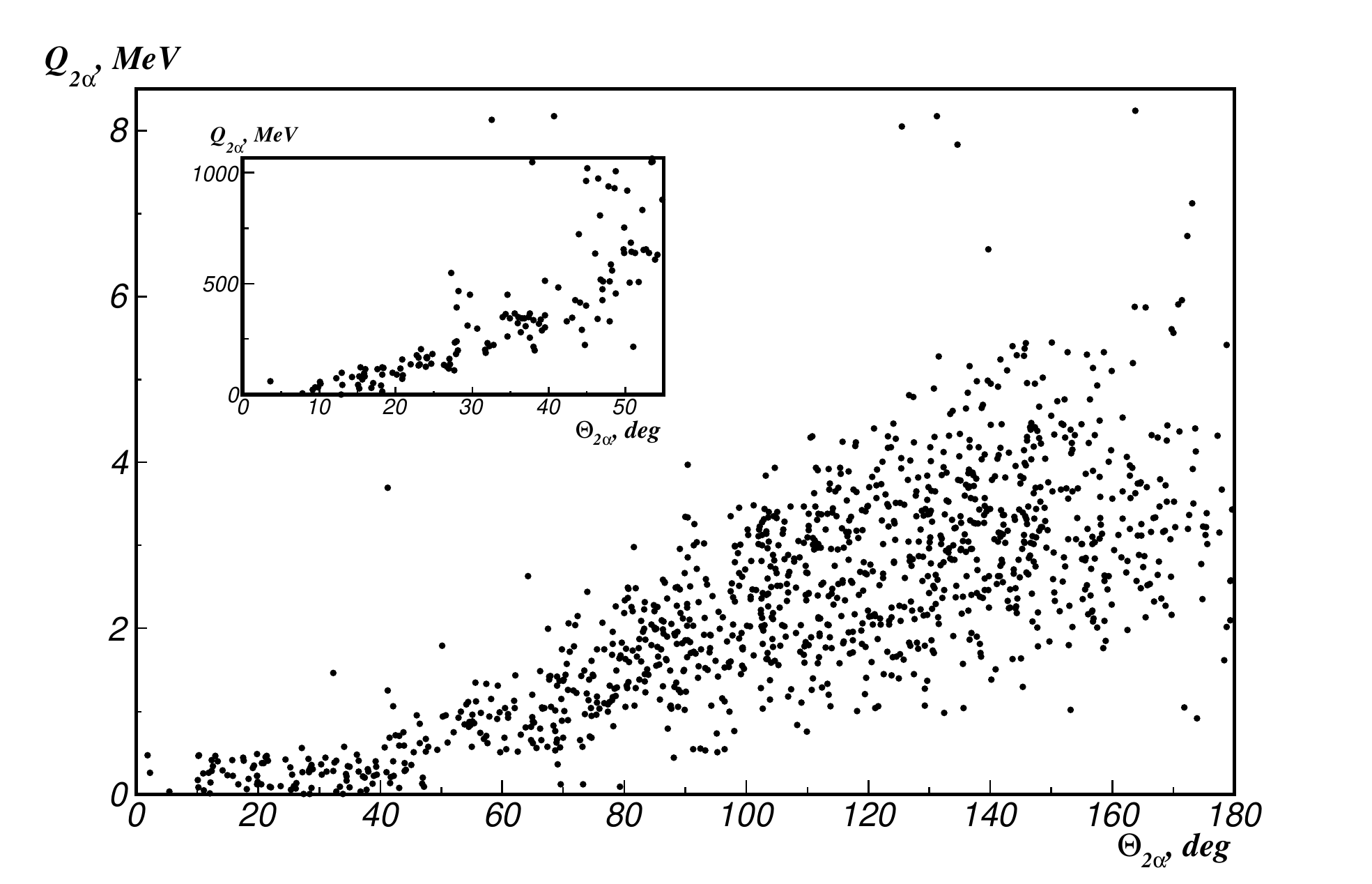}
\caption{Correlation over energy $Q_{2\alpha}$ and opening angles $\Theta_{2\alpha}$ in $\alpha$-particle pairs.}
\label{fig:4}
\end{figure}  
\begin{figure}
\includegraphics[width=0.45\textwidth]{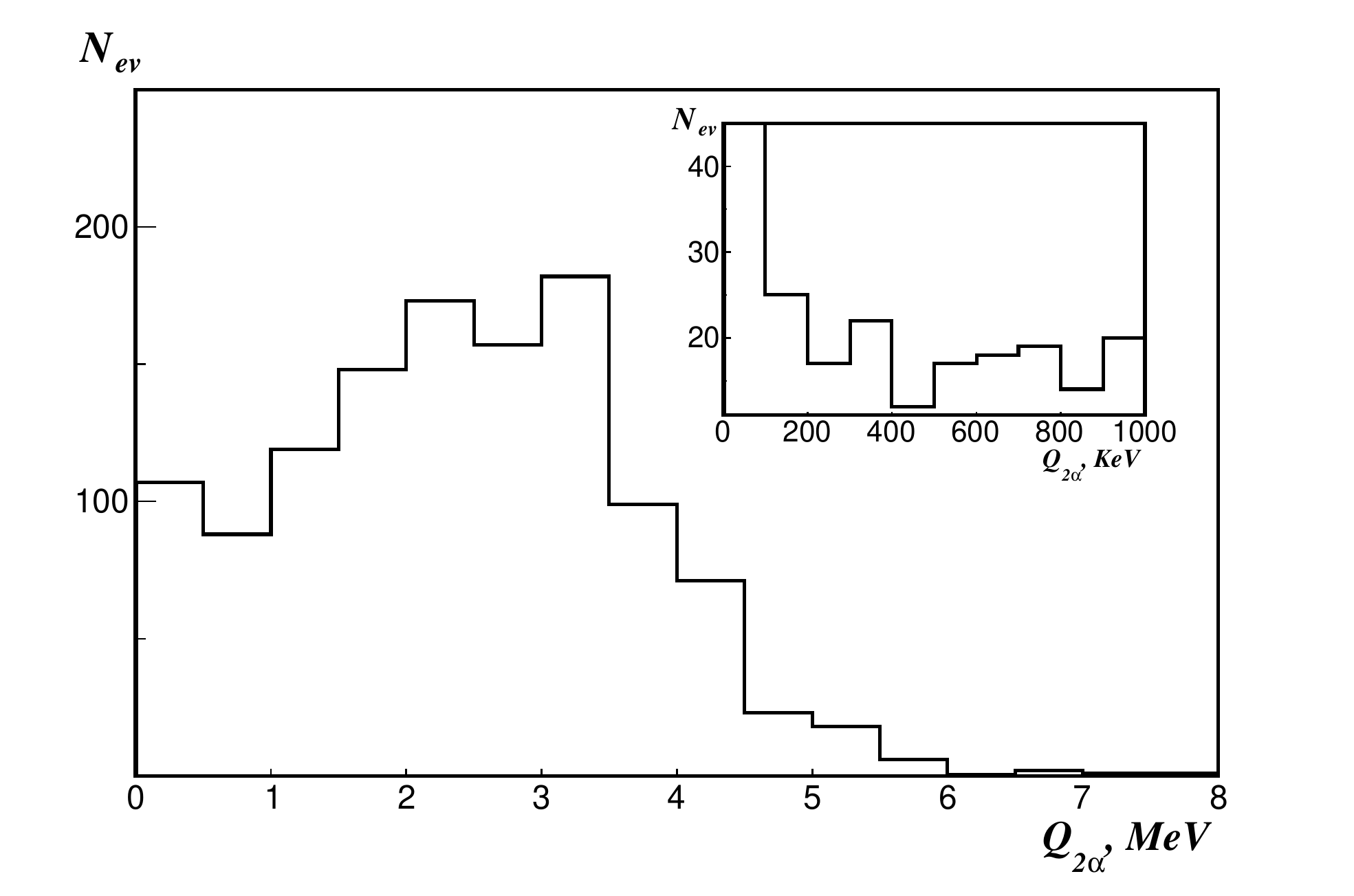}
\caption{Distribution pairs of $\alpha$-particles over energy $Q_{2\alpha}$.}
\label{fig:5}
\end{figure}

\indent Scanning of layers performed on microscopes MBI-9 was aimed at 3$\alpha$-disintegrations. In 400 events of 3$\alpha$-disintegration selected among the found 1200 ones measurements of angles relative to plane of a NTE layer and its surface as well as their lengths were at a KSM microscope performed for all $\alpha$-particle tracks. The only condition for the selection of the events was fullness of measure. Distribution over ranges of $\alpha$-particles L$_{\alpha}$ (Fig. 1) has an average value $<L_{\alpha}> = (5.8 \pm 0.2)$ $\mu m$ at RMS $(3.3 \pm 0.1)$ $\mu m$. This distribution has an asymmetric shape described by the Landau distribution. Directly associated with it the distribution over energy of $\alpha$-particles E$_{\alpha}$ (Fig. 2) defined by ranges $L_{\alpha}$ in the SRIM model [8] has an average value $<E_{\alpha}> = (1.86 \pm 0.05) MeV$ with RMS $(0.85 \pm 0.03) MeV$.\par
\indent Determination of angles and energy values versus ranges allows one to determine the energy $Q_{2\alpha}$ of pairs and triples $Q_{3\alpha}$ of $\alpha$-particles. Distribution over $Q_{3\alpha}$ (Fig. 3) is concentrated in the range of excitations of the $^{12}$C nucleus which is below thresholds of separation of nucleons. The used method does not resolve levels of $^{12}$C while the Hoyle state is not shown, as expected, for the reaction of $\alpha$-particle knocking out.\par
\indent Correlation over energy $Q_{2\alpha}$ and opening angles $\Theta_{2\alpha}$ in $\alpha$-particle pairs reveals features of the $^{8}$Be nucleus (Fig. 4). The region of large opening angles $\Theta_{2\alpha} > 90^{\circ}$ is corresponding to $Q_{2\alpha}$ of $^{8}$Be$_{2+}$, while $\Theta_{2\alpha} < 30^{\circ} - ^{8}$Be$_{g.s.}$. Distribution over $Q_{2\alpha}$ points to these states (Fig. 5). Its right side meets the shape expected from the decay through $^{8}$Be$_{2+}$. Condition $Q_{2\alpha} < 200 keV$ has allowed to allocate 56 decays $^{8}$Be$_{g.s.}$. For $^{8}$Be$_{g.s.}$ the total momentum distribution is rather narrow and characterized an average value of ($208 \pm 4) MeV/c$ with RMS $(30 \pm 3) MeV/c$.  Estimate of the average total momentum for 212 pairs of $\alpha$-particles which are the most appropriate to $^{8}$Be$_{2+}$ is $(130 \pm 3) MeV/c$ with RMS $(43 \pm 2) MeV/c$. Thus, the distribution over the total momentum for $^{8}$Be$_{2+}$ is much softer and relatively wider.\par
\indent Importance of the discussed structure is determined not only by interest to describe $^{12}$C$_{g.s.}$, but also the fact that it is the starting configuration for the reverse process of generating 3$\alpha$-particle ensembles in the Hoyle state. It is assumed that this state after $^{8}$Be$_{g.s.}$ is a Bose-Einstein condensate consisting of $\alpha$-particles with zero angular momentum [9]. Its identification in breakups of $^{12}$C allows one to advance to generation of condensate states of larger number of $\alpha$-particles. Fundamental aspect seems related to the fact that in order to recreate the condensate it is necessary to \lq\lq evacuate\rq\rq two hidden rotations $^{12}$C$_{g.s.}$. We note that in this respect the Coulomb dissociation of a nucleus on a heavy nucleus appears to be the most suitable process since few photon exchanges in it are possible.\par
\indent In general, these data indicate the presence of a superposition of states $0^{+}$ and $2^{+}$ of the nucleus $^{8}$Be in the ground state of $^{12}$C, and $^{8}$Be$_{2+}$ dominates. Deeper consideration of manifestation of $\alpha$-cluster structure of $^{12}$C in disintegrations caused by neutrons requires the use of the Dalitz plot and the theory of nuclear reactions. Presented measurements of the reaction $^{12}C(n,n')3\alpha$ and videos of these events are available on the BECQUEREL project website [10]. The authors express gratitude to S.~P.~Kharlamov (LPI) for a discussion of the results. This work was supported by grant from the Russian Foundation for Basic Research 12-02-00067 and grants Plenary representatives of Bulgaria, Romania and the Czech Republic in JINR.\par

\end{document}